# Imaging surface plasmon polaritons using proximal self-assembled InGaAs quantum dots

*Gregor Bracher, Konrad Schraml, Mäx Blauth, Jakob Wierzbowski, Nicolas Coca Lopez, Max Bichler, Kai Müller[†], Jonathan J. Finley, and Michael Kaniber[*]*

Walter Schottky Institut and Physik Department, Technische Universität München, Am Coulombwall 4, 85748 Garching, Germany

Nanosystems Initiative Munich, Schellingstraße 4, 80799 München, Germany

**We present optical investigations of hybrid plasmonic nanosystems consisting of lithographically defined plasmonic Au-waveguides or beamsplitters on GaAs substrates coupled to proximal self-assembled InGaAs quantum dots. We designed a sample structure that enabled us to precisely tune the distance between quantum dots and the sample surface during nano-fabrication and demonstrated that non-radiative processes do not play a major role for separations down to ~ $10\ nm$. A polarized laser beam focused on one end of the plasmonic nanostructure generates propagating surface plasmon polaritons that, in turn, create electron-hole pairs in the GaAs substrate during propagation. These free carriers are subsequently captured by the quantum dots ~ $25\ nm$ below the surface, giving rise to**

luminescence. The intensity of the spectrally integrated quantum dot luminescence is used to image the propagating plasmon modes. As the waveguide width reduces from 5 $\mu m$ to 1 $\mu m$, we clearly observe different plasmonic modes at the remote waveguide end, enabling their direct imaging in real space. This imaging technique is applied to a plasmonic beamsplitter facilitating the determination of the splitting ratio between the two beamsplitter output ports as the interaction length $L_i$ is varied. A splitting ratio of 50 ∶ 50 is observed for $L_i \sim 9 \pm 1$ $\mu m$ and 1 $\mu m$ wide waveguides for excitation energies close to the GaAs band edge. Our experimental findings are in good agreement with mode profile and finite difference time domain simulations for both waveguides and beamsplitters.

## I. INTRODUCTION

Recently, the coupling between electromagnetic fields and oscillations of the free electron gas at a metal-dielectric interface has attracted renewed interest [1]. In particular, modern nano-fabrication techniques such as electron beam lithography and focused ion beam milling have opened up the possibility to study and exploit surface plasmon polaritons in integrated geometries. Such systems have strong potential for applications in diverse research fields such as bio-sensing [2], graphene plasmonics [3], photovoltaics [4], nonlinear [5] and quantum optics [6]. Most recently, great progress has been made combining quantum emitters like molecules [7], colloidal quantum dots [8] or nitrogen-vacancy centers [9] with chemically synthesized nano-plasmonic structures. However, the desirable combination of such on-chip quantum light sources [10] with more sophisticated components such as plasmonic beam splitters and interferometers [11] or even integrated detectors [12], essentially forming an integrated plasmonic circuit, calls for an increased flexibility in designing and fabricating high-quality, sub-wavelength plasmonic structures [13]. Although, lithographically defined metallic nanostructures suffer from enhanced losses due to the polycrystalline nature of the metal [14], great advances have been reported by improving the optical properties of thermally evaporated metal via annealing [15] or using monocrystalline metal flakes in combination with focused ion beam milling [16]. A fully-integrated future plasmonic circuit ultimately requires on-chip light sources and, thus, the combination with semiconducting substrates promises the utilization of high-quality quantum emitters with excellent optical properties [17].

In this paper we present optical studies of lithographically defined plasmonic nanostructures on active GaAs substrates that contain near-surface self-assembled InGaAs/GaAs quantum dots [18]. For plasmonic waveguides [19] and beamsplitters, we demonstrate that the quantum dots can be used to image the propagating surface plasmon polariton modes [20] via their far-field emission, similar to well established fluorescence molecule imaging [21]. For waveguides we obtain evidence that different guided plasmonic modes [22] are supported with increasing waveguide width, each mode having a unique propagation length. Our findings are supported by finite difference time domain and mode profile simulations. Moreover, we apply our method to study plasmonic beamsplitters consisting of two adjacent waveguides with a width of $w_{WG} = 1\ \mu m$ and separated by a $g = 220\ nm$ gap over an interaction length $L_i$. We demonstrate that a common interaction length $L_i = 9 \pm 1\ \mu m$ between the two parallel waveguides is required to obtain an intensity splitting ratio $I_R/I_L$ close to 50:50 between the two output ports ($R$ and $L$) of the beamsplitter.

## II. SAMPLES AND EXPERIMENT

The samples studied were grown by solid source molecular beam epitaxy on a semi-insulating, [100] oriented GaAs substrate. Two samples (S1 and S2) were grown and nominally consisted of the following layer sequence; a $300\ nm$ GaAs buffer layer followed by a single layer of $In_xGa_{1-x}As$ quantum dots with a nominal indium content $x = 0.5$ grown at $560\ °C$ with a rate of $0.125\ Å/s$. These growth conditions resulted in an average quantum dot density estimated to be $\sim 140 \pm 10\ \mu m^{-2}$ for both samples. The samples were completed by overgrowing the quantum dot layer with a GaAs capping layer of $d_{cap,S1} = 120\ nm$ for S1 ($d_{cap,S2} = 25\ nm$ for S2), to make them optically active. By using optical lithography and wet chemical etching using a citric acid solution ($C_6H_8O_7\ /\ H_2O\ =\ 1\ g$ in $100\ ml$), we sequentially etched a series of steps into the GaAs capping layer [20]. This enabled us to control the distance between the quantum dot layer and the surface of the sample with a precision $\leq 4\ nm$, whilst maintaining a relatively smooth surface (RMS roughness $\leq 2\ nm$ - for details see supplementary material). Repeating this procedure, we obtain a staircase like sample as depicted schematically in Figure 1 (a). In total, ten different dot-surface separations ranging from $d = 120 \pm 1\ nm$ to $7 \pm 4\ nm$ were defined using this method on a single sample. Subsequently, electron beam lithography and lift-off processes were used to define rectangular gold (Au)-waveguides with widths $w_{WG} = 1\ \mu m - 5\ \mu m$ and lengths $l_{WG} = 10\ \mu m - 45\ \mu m$ on each step. The Au-waveguides were deposited using thermal evaporation and consisted of a nominally $t_{WG} = 0.1\ \mu m$ thick Au-layer [19]. In Figure 1 (b), we present a typical scanning electron microscopy image of (i) a plasmonic waveguide with dimensions $(w_{WG}, l_{WG}, t_{WG}) = (5\ \mu m, 35\ \mu m, 0.1\ \mu m)$ and (ii) a plasmonic beamsplitter with dimensions $(w_{WG}, L_i) = (1\mu m, 10\mu m)$, illustrating the high structural quality of the lithographically defined structures. Complementary atomic force microscopy measurements yield an average surface roughness of the Au-layer of $r_{rms} \sim 2\ nm$ (see supplementary material). At one end of the Au-waveguide a $450 \times 450\ nm^2$ square aperture was established to increase the in-coupling efficiency by scattering light incident perpendicular to the sample surface into propagating surface plasmon polariton modes. Depending on the width of the waveguides we obtain surface plasmon polariton propagation lengths $L_{SPP} = 10 - 40\ \mu m$, comparable to similar structures defined on unetched GaAs substrates [19].

The optical studies were performed using a low-temperature confocal microscope that provides diffraction limited performance with spot sizes $\leq 1\ \mu m$ and control of the polarization in both excitation and detection channels. For measurements, the samples were mounted on the cold finger of a continuous-flow helium cryostat and all experiments were conducted at nominal lattice temperature of $T = 15\,K$. For excitation, light from either a continuous-wave Helium-Neon laser ($E_{exc} = 1.959\,eV$), a pulsed laser diode ($E_{exc} = 1.884\,eV$, pulse width 50 $ps$, 40 $MHz$ repetition rate) or a tunable continuous-wave Ti:Sa laser ($E_{exc} = 1.512\,eV$) was focused onto the sample via an $100\times$ microscope objective (numerical aperture $NA = 0.80$). The emitted quantum dot luminescence was collected from an independently controllable detection area via the same objective, coupled into a single-mode fiber, dispersed using an 0.55 $m$ imaging spectrometer and detected with a liquid nitrogen cooled, Si-based charge coupled device (CCD) camera. In addition to this confocal detection scheme with separate positioning in excitation and detection channels, we also used a highly sensitive, Peltier-cooled CCD camera in combination with 900 $nm$ long pass filters to directly image the sample surface via the spectrally integrated quantum dot luminescence.

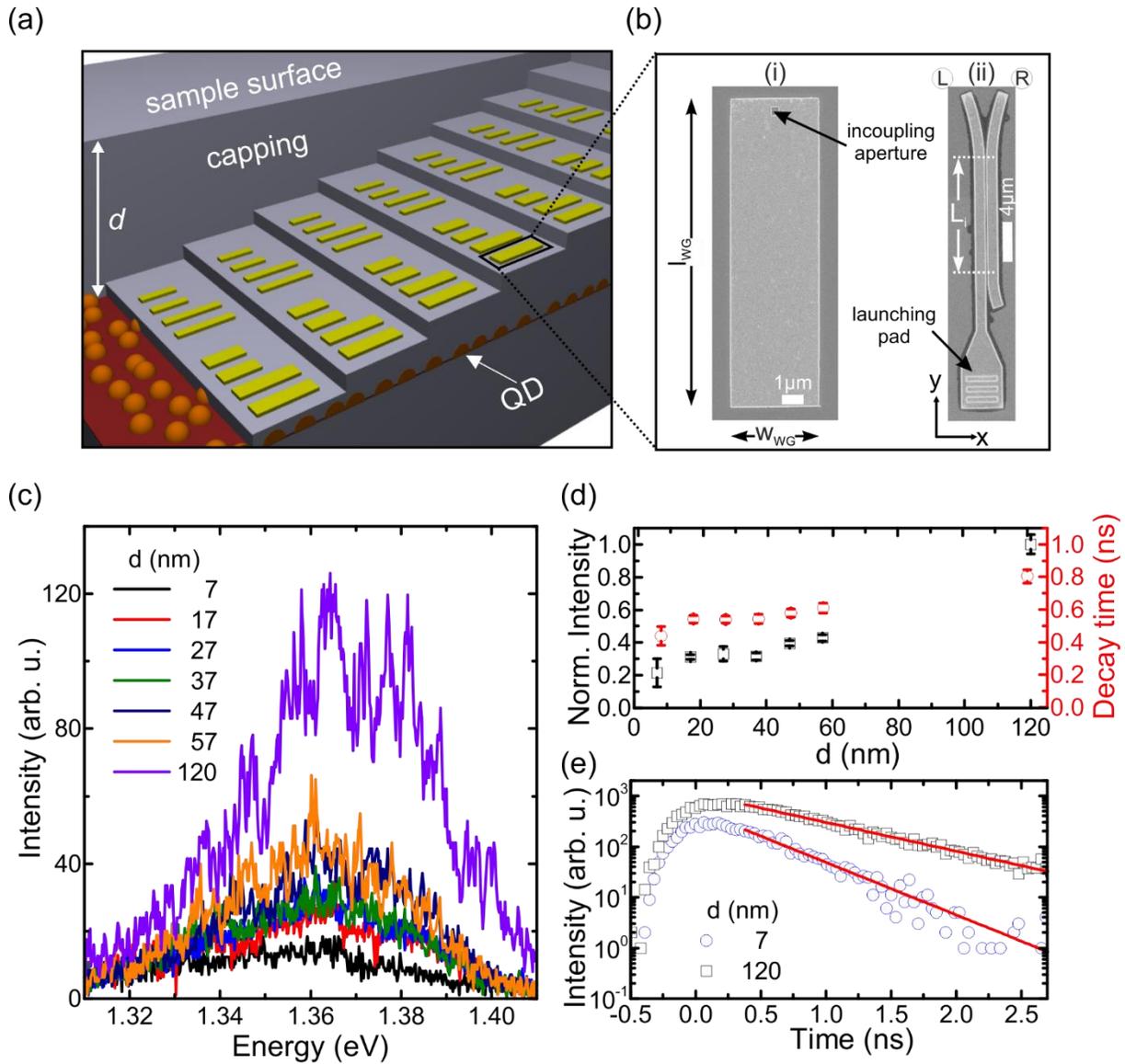

*Figure 1 - (a) Schematic sample layout. (b) Scanning electron microscope images of (i) a plasmonic waveguide and (ii) a plasmonic beamsplitter. (c) Typical quantum dot micro-photoluminescence spectra recorded for different quantum dot-surface separations d. (d) Time-integrated (black) and time-resolved (red) micro-photoluminescence as a function of d. (e) Time-resolved transients for $d = 7 \pm 4\ nm$ (blue circles) and $d = 120 \pm 1\ nm$ (black squares).*

III. RESULTS AND DISCUSSION

In Figure 1 (c) we present a series of micro-photoluminescence spectra recorded on quantum dots spatially displaced from the plasmonic nanostructures, $QD_{bulk}$, with different dot-surface separations $d$ ranging from $d = 120 \pm 1\,nm$ (upper spectrum, purple) to $7 \pm 4\,nm$ (lower spectrum, black). For these measurements we used a non-resonant laser excitation at $E_{exc} = 1.959\,eV$ and an excitation power density $P_{exc} \sim 10\,W/cm^2$. We observe a 60% decrease of the spectrally integrated quantum dot luminescence signal after the initial etching step as $d$ reduces from $120 \pm 1\,nm$ to $57 \pm 1\,nm$, shown by the black symbols in Figure 1 (d). We attribute this signal reduction to increased scattering due to the enhanced surface roughness introduced by the wet chemical etching as compared to the initially epitaxially flat wafer surface. Support for this identification is obtained from the observation that the signal then remains approximately constant down to $d = 17 \pm 3$ nm. Complementary time-resolved micro-photoluminescence spectroscopy performed on the different steps using pulsed excitation at $E_{exc} = 1.884\ eV$ (red symbols in Figure 1 (d)) reveals a shortening of the exciton spontaneous emission lifetime to $\tau_{57nm} = 0.6\ ns$ after the initial etching step when compared to the typical intrinsic lifetime $\tau_0 \sim 0.8\ ns$ as measured on the unetched region of the sample. In the range $17 \pm 3\ nm < d < 57 \pm 1\ nm$ the spontaneous emission decay time remains constant at $\sim 0.5 - 0.6\ ns$. Clear mono-exponential decay transients are observed for all dot-surface separations, as shown in Figure 1 (e) for $d = 120 \pm 1\ nm$ and $d = 7 \pm 4\ nm$ in black and blue, respectively. We conclude that although the quantum dot signal is partially quenched for reduced dot-surface separations and non-radiative processes become progressively more important for $d \leq 10\ nm$ [23], we observe measureable emission which can be used to probe the coupling of the quantum dots to near-surface plasmonic nanostructures.

We continue by performing luminescence spectroscopy on the plasmonic waveguides, whereby we optically excite via the square aperture and detect the resulting emission from the quantum dots surrounding the Au-waveguide. A typical spatially resolved image of the spectrally integrated quantum dot luminescence recorded around a $w_{WG} = 3\ \mu m$ wide plasmonic waveguide ($d = 27 \pm 2\ nm$) is presented in Figure 2 (a). We observe bright luminescence surrounding the plasmonic waveguide (dashed red line in Figure 2 (a)) that decays rapidly over lengthscales $< 1\ \mu m$, clearly indicating the strongly confined plasmonic field in the immediate vicinity of the waveguide. We interpret the quantum dot emission as arising from quantum dots that are locally excited by free charge carriers generated by the propagating surface plasmon polaritons close to the GaAs substrate. This proposed plasmon-mediated quantum dot excitation scheme is schematically depicted in Figure 2 (b). The linearly polarized laser is focused on the square aperture of the plasmonic waveguide, exciting propagating surface plasmon polaritons that are subsequently either scattered into the far-field (red curly arrows) or excite charge carriers in the near-by GaAs matrix (black arrow). Those charge carriers are efficiently captured by quantum dots $QD_{pl}$ in the immediate vicinity of the plasmonic waveguide (black dashed arrow), giving rise to luminescence between 1.30 $eV$ and 1.40 $eV$, red shifted from the excitation laser.

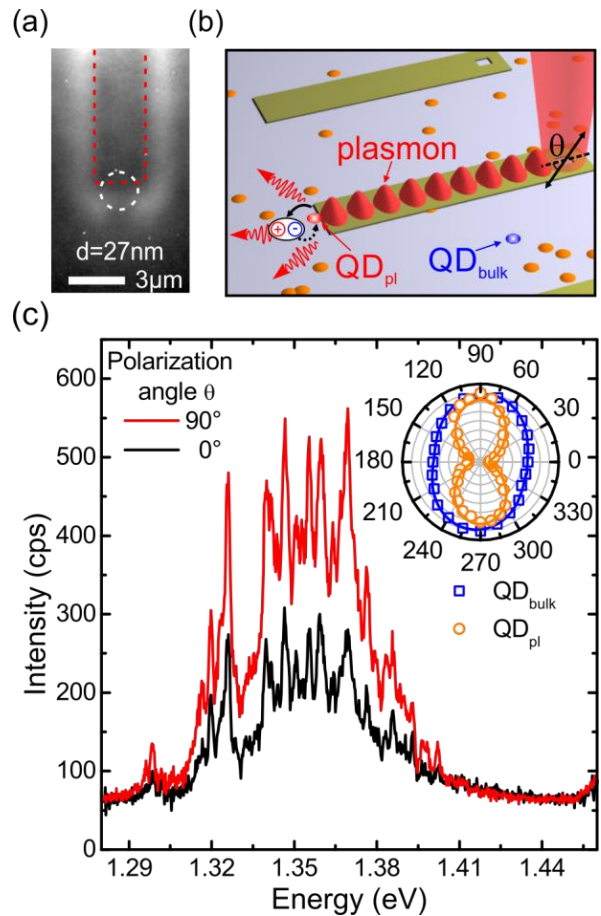

Figure 2 - (a) CCD camera measurement of the plasmonic waveguide imaged via the spectrally integrated QD signal. (b) Schematic excitation scheme of plasmon-excited quantum dots. (c) Polarization-resolved micro-photoluminescence spectra for $\theta = 0°$ (black) and $\theta = 90°$ (red). (Inset) Corresponding polar plot for plasmon-excited (orange) and bulk (blue) quantum dots.

In order to unambiguously associate the observed luminescence to proximal quantum dots excited via propagating plasmons, we probed the spectrum of the emission from the remote end of the plasmonic waveguide (dashed white circle in Figure 2 (a)) as a function of the excitation polarization as defined in Figure 2 (b). The quantum dot emission spectra are plotted in Figure 2 (c) for excitation polarization angles $\theta = 0°$ and $\theta = 90°$ in black and red, respectively. The observed spectra resemble typical quantum dot luminescence as recorded on the unpatterned region of the sample (c.f. Figure 1 (c)), giving rise to a broadband intensity distribution ($\Delta = 45 \pm 2\ meV$), decorated by sharp emission lines.

We observe a $\sim 2 \times$ stronger quantum dot emission from the waveguide end for a parallel ($\theta = 0°$) compared to a perpendicular ($\theta = 90°$) orientation of the excitation polarization as expected from the transverse magnetic character of the propagating surface plasmon polaritons [1]. The inset of Figure 2 (c) compares the integrated quantum dot photoluminescence intensity as a function of $\theta$ for plasmon-mediated excitation (orange) and confocal excitation (blue). For quantum dots excited via the surface plasmon polaritons, we obtain a degree of polarization $DoP_{pl} = \frac{I_{max} - I_{min}}{I_{max} + I_{min}} = 50 \pm 2\ \%$. In contrast, the signal from the reference quantum dots $QD_{bulk}$ using confocal spectroscopy is almost entirely unpolarized ($DoP_{bulk} = 15 \pm 2\ \%$). Further evidence for the plasmon-mediated generation of electron-hole pairs near the waveguide edges is given by the observation that this near-waveguide luminescence does not depend on $d$ and, therefore, is not related to plasmonic near-field excitation of quantum dots (for details see supplementary material).

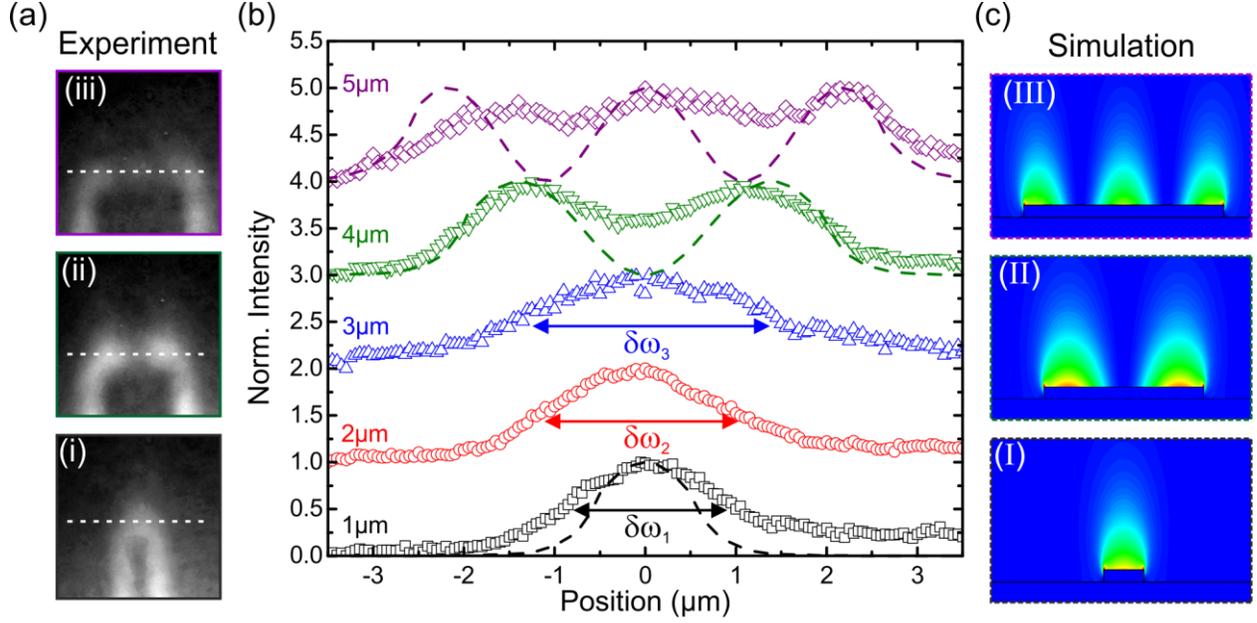

*Figure 3 - (a) Spatially resolved quantum dot photoluminescence images for waveguide width (i) $w_{WG} = 1\ \mu m$, (ii) $w_{WG} = 4\ \mu m$ and (iii) $w_{WG} = 5\ \mu m$. (b) Extracted quantum dot intensity as a function of position at the remote waveguide end for waveguide widths $w_{WG} = 1\ \mu m$ (bottom) to $5\ \mu m$ (top). (c) Corresponding simulation of the plasmon mode for (I) $w_{WG} = 1\ \mu m$, (II) $w_{WG} = 4\ \mu m$ and (III) $w_{WG} = 5\ \mu m$.*

Those observations unambiguously identify the emission surrounding the plasmonic waveguide as arising from quantum dots that are optically excited via propagating surface plasmon polaritons [19].

The plasmon-mediated excitation of proximal quantum dots can be used to image propagating surface plasmon polaritons via their emission into the far-field [21]. Hereby, we imaged the spectrally integrated quantum dot emission from the waveguide end for different $w_{WG}$ and $l_{WG}$. Figure 3 (a) shows typical spatially resolved CCD camera images of the quantum dot emission for waveguide width $w_{WG} = 1\ \mu m$, $4\ \mu m$ and $5\ \mu m$ in panels (i) - (iii), respectively. Plotting the cross-sections of the quantum dot emission along the end of the waveguides (indicated by the dashed lines), we obtain spatial intensity profiles as shown in Figure 3 (b). For $w_{WG} = 1\ \mu m$ (black curve) we observe a single peak centered at the waveguide midpoint with a full width of half maximum (FWHM) $\delta w_1 = 1.80 \pm 0.05\ \mu m$, reflecting the lateral waveguide dimension[1]. For increasing waveguide width we still observe a single peak, however, with increased FWHM of $\delta w_2 = 2.06 \pm 0.04\ \mu m$ and $\delta w_3 = 2.78 \pm 0.07\ \mu m$ for $w_{WG} = 2\ \mu m$ (red curve) and $w_{WG} = 3\ \mu m$ (blue curve), respectively. In strong contrast, wider waveguides having $w_{WG} = 4\ \mu m$ (green curve) and $w_{WG} = 5\ \mu m$ (purple curve) clearly exhibit two and three maxima, respectively, which are attributed to higher order plasmonic modes [22] [24]. In order to support this hypothesis, we performed mode profile simulations [25] of the plasmonic modes generated at the Au-air interface for $w_{WG} = 1\ \mu m$, $4\ \mu m$ and $5\ \mu m$ and the results are plotted in panels (I) - (III) of Figure 3 (c), respectively. We observe an excellent qualitative agreement regarding the number of plasmonic modes, yielding one, two and three antinodes for waveguides with $w_{WG} = 1\ \mu m$, $4\ \mu m$ and $5\ \mu m$, respectively. Moreover, we obtain also good agreement for the spatial position of the mode maxima as shown by the simulation data, presented as the dashed curves in Figure 3 (b) for $w_{WG} = 1\ \mu m$ (black curve), $4\ \mu m$ (green curve) and $5\ \mu m$ (purple curve). We note here that we observe in our experiments a superposition of several modes for wider waveguides ($w_{WG} > 3\ \mu m$), resulting in a reduced contrast between adjacent maxima and minima as compared to the simulation of the individual modes (dashed curves in Figure 3 (b)). This assertion is further supported by photoluminescence experiments performed on $w_{WG} = 5\ \mu m$ waveguides exhibiting different lengths $l_{WG}$ which clearly show that the number of maxima observed at the waveguide terminations and, thus, the number modes reduces with increasing $l_{WG}$ due to the smaller propagation length of the higher order modes (for experimental data see supplementary material) [24]. Our experimental findings are in agreement with near-field optical studies on similar metallic waveguides using photon scanning tunneling spectroscopy [22] [24].

---

[1] The deviation between the actual waveguide width $w_{WG,SEM} = 1\ \mu m$ extracted from scanning electron microscopy images and the measured mode confinement arises from the convolution with the limited detection resolution in those non-confocal CCD camera measurements.

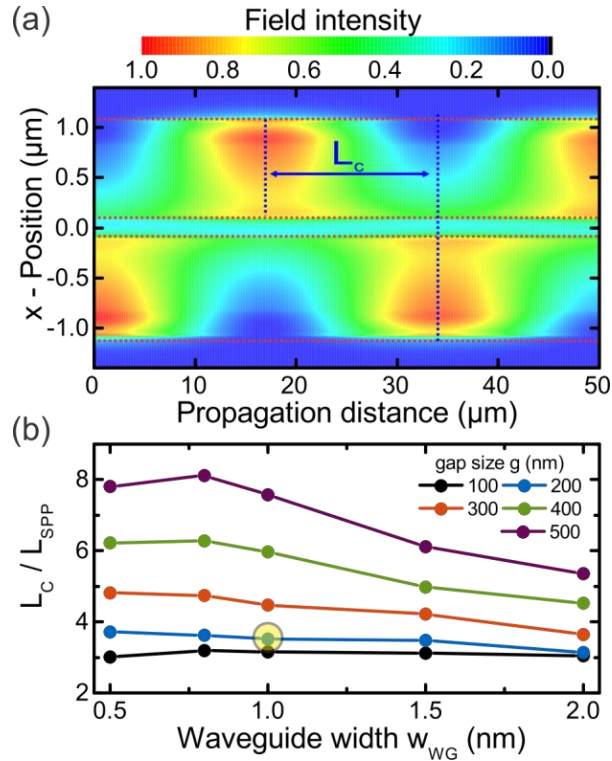

*Figure 4 - (a) Simulation of the plasmonic mode profile as a function of propagation distance. (b) Ratio $L_C/L_{SPP}$ as a function of $w_{WG}$ for gap sizes $g = 100\ nm\ to\ 500\ nm$.*

To further emphasize the imaging capability of our hybrid system, we realized a plasmonic beamsplitter[2] using two coplanar Au-waveguides separated by a gap $g$ along a well-defined interaction length $L_i$ [26]. A typical scanning electron microscopy image of such a structure is shown in panel (ii) of Figure 1 (b). We present in Figure 4 (a) a finite difference time domain simulation of the plasmonic mode intensity profile (encoded in color) as a function of x-position and propagation distance along the y-direction for $w_{WG} = 1\ \mu m$ and a fixed spacing of $g = 220\ nm$.[3] In this simulation the surface plasmon polariton is launched in the lower waveguide ($x < 0$) and the maximum of the mode intensity oscillates between the lower and the upper waveguide with increasing plasmon propagation distance. The spacing between two consecutive maxima is defined as the coupling length $L_C$ needed to transfer the electromagnetic energy from the lower to the upper waveguide. In Figure 4 (b) we present the calculated $L_C$ normalized to the plasmon propagation length $L_{SPP}$, as a function of $w_{WG}$ and $g$. Firstly, we find that for a constant $w_{WG}$ the ratio $L_C/L_{SPP}$ strongly increases with increasing $g$, reflecting the less efficient exchange of the electromagnetic field between the two adjacent waveguides for larger gap sizes. Secondly, we observe for $g \geq 200\ nm$ an increase of $L_C/L_{SPP}$ with decreasing $w_{WG}$, due to the strongly decreasing $L_{SPP}$, which is dominating the ratio $L_C/L_{SPP}$ for small $w_{WG}$.

---

[2] Those experiments have been realized on sample S2.

We experimentally investigated beamsplitters with $w_{WG} = 1\ \mu m$ and $g = 220\ nm$ (c.f. yellow circle in Figure 4 (b)). Propagating surface plasmon polaritons are excited via the launching pad (c.f. scanning electron microscopy image in Figure 1 (b)) and plasmon propagation is imaged via the spatially resolved quantum dot photoluminescence as shown in panel (i), (ii) and (iii) of Figure 5 (a) for $L_i = 3\ \mu m$, $7\ \mu m$ and $10\ \mu m$, respectively.

---

[3] For the sake of presentation we normalized the mode intensity to the theoretically obtained propagation losses of the surface plasmon polaritons in this simulation study.

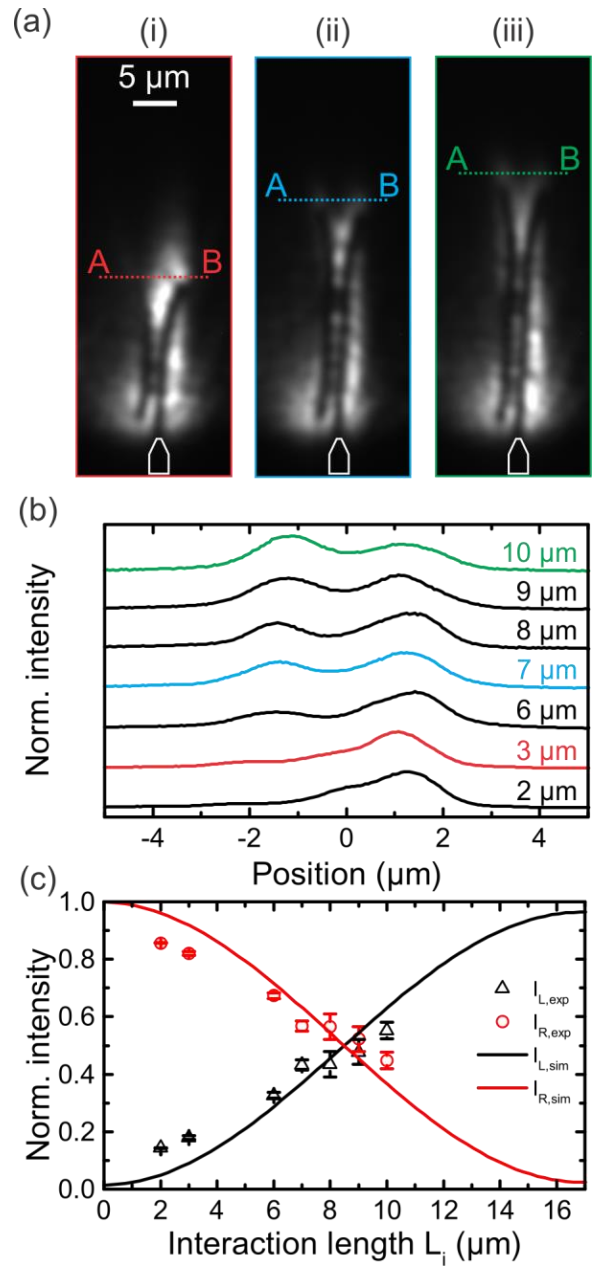

*Figure 5 - (a) Spatially resolved quantum dot photoluminescence images for plasmonic beamsplitters with interaction length (i) $L_i = 3$ μm, (ii) $L_i = 7$ μm and (iii) $L_i = 10$ μm. (b) Cross-sectional intensity profiles for $L_i = 2$ μm (lower spectrum) to $10$ μm (upper spectrum). (c) Experimentally determined normalized output intensity (symbols) and comparison with simulation (solid lines).*

We clearly observe the waveguides as dark regions, whilst the outline of the structures can be visualized in Figure 5 (a) via the quantum dot luminescence. In Figure 5 (b), we present cross sections of the quantum dot intensity at the remote end of the plasmonic beamsplitters along the lines AB (c.f. dashed lines in Figure 5 (a)) for $L_i = 2\ \mu m$ (lower curve) to $L_i = 10\ \mu m$ (upper curve). Two maxima corresponding to the beamsplitter output ports are clearly seen in Figure 5 (b). We observe that a larger fraction of the intensity is detected via the exit $R$ for $L_i = 2\ \mu m$ (c.f. Figure 1 (b)). With increasing $L_i$ the intensity ratio between $R$ and $L$ changes and approaches approximately $50:50$ for $L_i \sim 9 \pm 1\ \mu m$. Further increase of $L_i$ to $10\ \mu m$ results in an increased intensity detected via the exit $L$. These measurements clearly demonstrate the ability to adjust the splitting ratio $I_R/I_L$ via the careful selection of the interaction length $L_i$ during fabrication.

In Figure 5 (c) we plot the relative intensities of the output in the R and L ports as a function of $L_i$ for our experimental results (symbols) and compare with finite-difference time domain simulations (solid lines). We obtain a good quantitative agreement between experiment and simulation, yielding a $50:50$ intensity splitting at $L_{i,50:50}^{ex} = 9 \pm 1\ \mu m$ ($L_{i,50:50}^{sim} = 8.5 \pm 0.1\ \mu m$).

Our results demonstrate the potential of the proposed plasmonic beamsplitter approach and show that the combination with near-surface quantum dots exhibits a powerful tool for surface plasmon imaging. However, we note that the limited $L_{SPP}$ for such plasmonic beamsplitters results in high $L_C/L_{SPP}$-ratios of $> 3$ for realistic fabrication parameters such as $w_{WG} = 1\ \mu m$ and $g = 220\ nm$. Such high $L_C/L_{SPP}$-ratios indicate that the minimum interaction length $L_{i,min} = L_C/2$ to obtain a splitting ratios $I_R/I_L = 50:50$ is much longer than the typical $L_{SPP}$ [19], limiting the potential of this beamsplitter design in its current form for real-world applications. However, we believe that such directional couplers could be further improved by employing chemically synthesized monocrystalline metal flakes [16] or epitaxially grown films [27] in combination with slightly varied plasmonic structures such as for example dielectric-loaded surface plasmon polariton waveguides [28]. The latter has recently been used to demonstrate unambiguous signature of two-plasmon quantum interference [29].

IV. SUMMARY AND CONCLUSIONS

In conclusion, we designed and fabricated an optically active semiconductor chip with well-defined dot-surface separations and demonstrated the usability of near-surface semiconductor quantum dots to image surface plasmon polaritons in lithographically defined waveguides of $w_{WG} = 1\ \mu m$ to $5\ \mu m$ and plasmonic beamsplitters consisting of two parallel waveguides with $w_{WG} = 1\ \mu m$ and $g = 220\ nm$. We observe one single mode for $w_{WG} = 1\ \mu m$ to $3\ \mu m$ whilst two and three modes are obtained for $w_{WG} = 4\ \mu m$ and $w_{WG} = 5\ \mu m$, respectively, in excellent agreement with simulations and literature [22] [24]. For the plasmonic beamsplitter we demonstrate that the intensity splitting ratio $I_R/I_L$ between the two output ports $R$ and $L$ can accurately be controlled via the common interaction length $L_i$ of the two adjacent waveguides forming the beamsplitter. For $L_i = 9 \pm 1\ \mu m$ we obtain $I_R/I_L \sim 50:50$, thus, enabling us to engineer the exact properties via this particular beamsplitter design.

ASSOCIATED CONTENT

Further details regarding the fabrication process, the experimental technique and the plasmonic mode profile simulations can be found in the supplementary material.

AUTHOR INFORMATION

**Corresponding Author**

*Electronic mail: Michael.Kaniber@wsi.tum.de

**Present Addresses**

† Kai Müller, Ginzton Laboratory, Stanford University, Stanford, California 94305, USA.

ACKNOWLEDGMENT

We acknowledge financial support of the DFG via the SFB 631, Teilprojekt B3, the German excellence Initiative via Nanosystems Initiative Munich, and FP-7 of the European Union via SOLID. The authors gratefully acknowledge the support of the TUM Graduate School's Faculty Graduate Center Physik at the Technische Universität München and of the International Graduate School for Science and Engineering (IGSSE).